\documentclass[aps,preprintnumbers,prl,twocolumn,superscriptaddress]{revtex4}

\usepackage{slashed}
\usepackage{graphicx,color}
\usepackage{epsfig}
\usepackage{subfigure}
\usepackage{epsfig}
\usepackage{dcolumn}
\usepackage{bm}
\usepackage{color}

\begin{document}

\preprint{ACFI-T16-01}

\title{The Diphoton Excess from an Exceptional Supersymmetric Standard Model}

\author{Wei Chao}
\email{chao@physics.umass.edu}

 \affiliation{ Amherst Center for Fundamental Interactions, Department of Physics, University of Massachusetts-Amherst
Amherst, MA 01003, United States }

\vspace{3cm}

\begin{abstract}

In this paper we explain the diphoton excess in the invariant mass $M \approx 750~{\rm GeV}$, claimed by the ATLAS and CMS collaborations at the run-2 LHC, as the signal of a scalar singlet in a string-inspired exceptional supersymmetric standard model (ESSM).  
The scalar singlet might play a rule  in the spontaneous breaking of the $U(1)^\prime$ gauge symmetry of the ESSM and  couples to diphoton and/or gluon pair with the help of exotic quarks and Higgs-like  supermultiplets, which are contained in the fundamental representation of the $E_6$ group. 
The model might give rise to a large enough production cross section at the LHC but can hardly fit with the wide width of the resonance except in the strong couple regime. 
\end{abstract}

\maketitle

\section{Introduction}
The discovery of the Higgs boson at the LHC has completed the story of the standard model (SM) of particle physics, which is in spectacular agreement with almost all experiments.   
But the SM can  not be the final theory because of many unsolved problems in the framework of the SM, such as neutrino masses, dark matter and baryon asymmetry (BAU).  
Recently, both the ATLAS~\cite{ATLAS13} and CMS~\cite{CMS13} collaborations have observed a resonance of invariant mass $750$ GeV at the run-2 LHC in the diphoton channel. 
The significance is $3.6\sigma$ for the ATLAS and $2.6\sigma$ for the CMS. 
If confirmed, it will be the first evidence of new physics beyond the SM from collider experiments. 
There are a bunch of interesting interpretations~\cite{Angelescu:2015uiz,Backovic:2015fnp,Bellazzini:2015nxw,Buttazzo:2015txu,DiChiara:2015vdm,Ellis:2015oso,Franceschini:2015kwy,Gupta:2015zzs,Harigaya:2015ezk,Higaki:2015jag,Knapen:2015dap,Low:2015qep,Mambrini:2015wyu,McDermott:2015sck,Molinaro:2015cwg,Nakai:2015ptz,Petersson:2015mkr,Pilaftsis:2015ycr,Dutta:2015wqh,Cao:2015pto,Matsuzaki:2015che,Kobakhidze:2015ldh,Martinez:2015kmn,Cox:2015ckc,Becirevic:2015fmu,No:2015bsn,Demidov:2015zqn,Chao:2015ttq,Fichet:2015vvy,Curtin:2015jcv,Bian:2015kjt,Chakrabortty:2015hff,Agrawal:2015dbf,Csaki:2015vek,Falkowski:2015swt,Aloni:2015mxa,Bai:2015nbs,Chao:2015nsm,Han:2015cty,xxx,Ibanez:2015uok} of the result as manifestations of various new physics. 
To figure out which new physics is the best explanation of the $750$ GeV diphoton excess, one needs more data and a higher luminosity.

The minimal supersymmetric standard model (MSSM) is one of the most promising new physics beyond the SM, because it provides solutions to the hierarchy problem, dark matter and gauge couplings unification.  
But the MSSM suffers from the $\mu$ problem:  if $\mu$, the coupling of the bilinear term $ H_u H_d$ in the superpotential, equals to the planck mass, then electroweak symmetry breaking does not occur. 
Besides, there is conflict between the $125$ GeV Higgs boson and the strongly first order electroweak phase transition~\cite{Katz:2015uja}, a necessary requirement of generating the BAU via the electroweak baryogenesis mechanism.
One another compelling theoretical argument of new physics is the unification of the SM gauge interactions with the gravity.  
An underlined theory is the supergraivty, which might be an effective low energy limit of the ten dimensional heterotic superstring theory based on the $E_8^{} \times E_8^\prime$~\cite{Green:1987mn}.
The compactification of the extra dimensions leads to the breakdown of $E_8$ to $E_6$ subgroup,  leaving $E_8^\prime$ as the hidden sector that plays the key rule in the spontaneous breaking of the supergravity. 
Thus one naturally arrive at a string inspired $E_6$ supersymmetry model, referred to as ESSM. 
At the  GUT scale, $E_6$ can be broken directly into $SU(3)_c\times SU(2)_L \times U(1)_Y \times U(1)_\psi \times U(1)_\chi$ via the Hosotani mechanism~\cite{Hosotani:1983vn}, where $U(1)_\psi$ and $U(1)_\chi$ gauge symmetries are defined as~\cite{King:2005jy} $E_6\to SO(10)\times U(1)_\psi$ and $SO(10)\to SU(5) \times U(1)_\chi$. 
Two extra gauge symmetries $U(1)_\psi \times U(1)_\chi$ can be reduced to one gauge symmetry $U(1)^\prime$: $U(1)^\prime \equiv \sin \theta U(1)_\psi+ \cos \theta U(1)_\chi$.  
In this case, the $\mu$ term is automatically forbidden by the gauge symmetry. 
It can only be generated dynamically by introducing  a SM singlet $S$, coupling to Higgs superfields, $\lambda S  H_u H_d$, whose vacuum expectation value (VEV) breaks the $U(1)^\prime$ spontaneously. 
Besides  it is easier to get a heavy SM Higgs boson in ESSM comparing with the MSSM case~\cite{Athron:2012sq}. 
These two advantages make ESSM more attractive than MSSM.
The $E_6$ inspired models with $U(1)^\prime$ have been extensively studied in Refs.~\cite{King:2005jy,Athron:2009bs,Athron:2012sq,Athron:2011wu,Langacker:1998tc,Suematsu:1994qm,Keith:1997zb,Daikoku:2000ep,Kang:2004pp,Chao:2014hya}, of which the scenario $U(1)_N$,  under which right-handed neutrinos have zero charge, was first highlighted in Ref.~\cite{King:2005jy}.  
It corresponds to  $\sin\theta =\sqrt{15}/4$. 
The advantage of this scenario is that active neutrino masses can be naturally generated by the canonical type-I seesaw mechanism.

In this paper we interpret the 750 GeV diphoton excess observed at the run-2 LHC as one of the scalar singlets $S$ that triggers the spontaneous breaking of the $U(1)_N$ gauge symmetry in the ESSM. 
$S$ might be produced at the LHC through the gluon fusion with exotic quarks running in the loop, which are components of the fundamental $\mathbf{27}$ representation of the $E_6$. 
The decay of $S$ into diphoton arises at the one-loop level, mediated by the exotic quarks and Higgs-like  supermultiplets. 
We find that the observed cross section can be explained in this model without conflicting with any experimental observations.
The wide width of $S$ also might be derived in this model, but only in the strong couple regime. 
Compared with other well-studied ESSM scenarios, our model is different in breaking the $U(1)_N$ with two scalar singlets. 
The mass spectrum of the CP-even scalars, neutralinos and exotic quarks in this model are generated. 

The remaining of the paper is organized as follows: In section II we describe our model briefly.  In section III we calculate mass spectrums.  Section IV is devoted to the interpretation of the diphoton excess in this model. The last part is concluding remarks.

\section{The model}

In this section, we give a brief introduction to the ESSM studied in this paper, which is a modified version of the one proposed in Ref.~\cite{King:2005jy}. 
The model contains an additional $U(1)_N$ gauge symmetry, and three complete fundamental $\mathbf{27}$ representations of $E_6$ in the particle content so as to spontaneously cancel anomalies of the $U(1)_N$. 
The decomposition of the model under the  $SU(5)\times U(1)_N$ takes the fom
\begin{eqnarray}
27&\to& (10,~{1\over \sqrt{40}}) \oplus (5^*,~{2\over \sqrt{40}}) \oplus (5^*,~-{3\over \sqrt{40}}) \nonumber \\
&&\oplus (5,~-{2\over \sqrt{40}}) \oplus (1,~{5\over \sqrt{40}}) \oplus (1,~0),
\end{eqnarray}
where the first and second terms in the brackets are the $SU(5)$ representation and $U(1)_N$ charge respectively.
The left-handed quark doublets, right-handed up-type quarks and right-handed charged leptons are assigned to the $(10,~1/\sqrt{40})$. 
The left-handed lepton doublets and right-handed down-type quarks are assigned to the $(5^*,~2/\sqrt{40})$.  
Two Higgs doublets as well as exotic quarks $D(\bar D) $, transforming as $(3^{(*)},~1,~-1/3,-3(-2))$ under the SM gauge group, are assigned to the $(5^*,~-{3/\sqrt{40}})\oplus(5,~-2/\sqrt{40})$. 
$(1,~5/\sqrt{40})$ corresponds to scalar singlets. Right-handed neutrinos are associated with $(1,~0)$. 
The renormalizable superpotential originates from the the $\mathbf{27}\times\mathbf{27}\times\mathbf{27}$ decomposition of the $E_6$ fundamental representation. 
It can be written as $W\equiv W_0 + W_1$, with~\cite{King:2005jy}
\begin{eqnarray}
W_0^{} &=& \lambda_{ijk}^{} S^i H_u^j H_d^k + \kappa_{ijk}^{} S^i D^j \bar D^k+W_{\rm MSSM}  \; , \label{superpot} \\
W_1^{} &=& g_{ijk}^{Q} D^i (Q^j Q^k) + g_{ijk}^q \bar D^i d^{cj} u^{ck}  \; , \label{diquark} 
\end{eqnarray}
where $W_{\rm MSSM}$ is the superpotential of the MSSM except the $\mu $ term.
We have taken $D_i$ as diquark with baryon number $2/3$ when write down eq. (\ref{diquark}). 
The superpotential contains too many irrelevant couplings, which can be omitted by applying a slightly broken $Z_2^H$ symmetry to the superpotential. 
Only  $S^2, S^3, H_u^3, H_d^3$ are even, while all other superfields are odd under the $Z_2^H $ symmetry.  
Further assuming the Yukawa coupling of $S^\alpha$ has flavor diagonal structure, the first two terms in the eq. (\ref{superpot}) can be simplified as 
\begin{eqnarray}
W_0 \ni \sum_{\alpha=2}^3 \sum_{i=1}^3 S^\alpha (\lambda_{\alpha  i} H_u^i H_d^i + \kappa_{\alpha i} D^i \bar D^i ) \; . \label{exotic}
\end{eqnarray}
Actually one does not need to worry about the flavor changing neutral current problems induced by the Yukawa interactions of $S^2$ and $S^3$, because they only couple to beyond SM particles and the Higgs boson. 
Comparing with the conventional ESSM case, we have two scalar singlets triggering the spontaneous breaking of the $U(1)_N$, which is the typical character of this model.  The particle contents of the model is listed in table.~\ref{pcontent}. 
\begin{table}[htbp]
\centering
\begin{tabular}{cccccccccccc}
\hline\hline  Particles & Q & $u^c$ & $d^c$ & $\ell$ & $E^c $ & $N^c$ & S & $H_u$ & $H_d$ & D & $\bar D$\\
$\sqrt{40} Q_N$& 1& 1&2&2&1&0&5&-2&-3&-2&-3 \\
\hline
\hline 
\end{tabular}
\caption{ Particle contents of the ESSM and their charges under the $U(1)_N^{}$.  }\label{pcontent}
\end{table}

\section{Mass spectrum}

The Higgs sector responsible for the $U(1)_N^{}$ and the electroweak symmetries breaking  includes two Higgs doublets $H_u^3$, $H_d^3$ as well as two scalar singlets $S_2$ and $S_3$. 
We assume there is no CP violation in the Higgs potential and set $\lambda \equiv \lambda_{33}^{} $, $  \tilde \lambda\equiv \lambda_{23}^{}  $, $H_u\equiv H_u^3$, $H_d^{} \equiv H_d^3$, $S\equiv S_3$ and $S^\prime \equiv S_2$ for simplification. 
The general scalar potential takes the form $V=V_F+V_D+V_{s}$, where
\begin{eqnarray}
V_F^{} &=& (\lambda^2 |S |^2 + \tilde \lambda^2  |S^\prime|^2 )(|H_u^{} |^2+|H_d^{} |^2 )+ \nonumber \\
&&\lambda\tilde \lambda(S^\dagger S^\prime + S^{\prime \dagger } S ) ( |H_u^{}|^2+ |H_d^{}|^2 ) + \nonumber \\ 
&&(\lambda^2  + \tilde\lambda^2)|H_u H_d|^2. \\
V_D^{} &=&  {g_2^2\over 2 } |H_u^\dagger H_d^{}|^2 + {g_1^2 +g_2^2 \over 8}  (|H_u|^2-|H_d|^2)^2 + \nonumber \\
&& {g^{\prime 2}\over 2 } [Q_{u}^2 |H_u|^2 + Q_d^2| H_d|^2 + Q_s^2 (|S|^2 + |S^\prime|^2)]^2. \\
V_{s}&=& M_S^2 |S|^2 + M_{S^\prime}^2 |S^\prime|^2  + M_{u}^2 |H_u|^2 + M_{d}^2 |H_d|^2  \nonumber \\
&&+\sqrt{2} ( \lambda A_\lambda S + \tilde\lambda A_{\tilde \lambda } S^\prime ) H_u H_d + {\rm h.c.}.
\end{eqnarray}
We have neglected loop corrections to the potential.
The VEVs of the Higgs fields are
\begin{eqnarray}
&&\langle H_u \rangle =\left(  \matrix{0 \cr {v_u \over \sqrt{2}}}\right) \;  ,\hspace{0.5cm}
\langle S\rangle = {v_s \over \sqrt{2}}  \; , \\
&&\langle H_d \rangle =\left(  \matrix{{v_d \over \sqrt{2}} \cr 0}\right)\; , \hspace{0.5cm}
\langle S^\prime \rangle = {v_{s^\prime} \over \sqrt{2}} \; .
\end{eqnarray}
The tadpole conditions can be written as
\begin{eqnarray}
0&=&(\lambda v_s + \tilde \lambda v_{s^\prime})^2 + ( \lambda^2 + \tilde \lambda^2 ) v_d^2  + {\bar g^2 \over 4 } (v_u^2-v_d^2) + 2 M_u^2\nonumber \\
&&+{g^\prime}^2 Q_u^2 \sum_i Q_i^2 v_i^2 - 2 {v_d \over v_u } (\lambda A_\lambda v_s + \lambda A_\lambda v_{s^\prime}) \\
0&=&(\lambda v_s + \tilde \lambda v_{s^\prime})^2 + ( \lambda^2 + \tilde \lambda^2 ) v_u^2  + {\bar g^2 \over 4 } (v_d^2-v_u^2) + 2 M_d^2\nonumber \\
&&+{g^\prime}^2 Q_d^2 \sum_i Q_i^2 v_i^2 - 2 {v_u \over v_d } (\lambda A_\lambda v_s + \lambda A_\lambda v_{s^\prime}) \\
0&=& \lambda (\lambda v_s + \tilde \lambda v_{s^\prime}){v_u^2 + v_d^2 \over 2 }+ {{g^\prime}^2 Q_s^2 \over 2 }  \sum_i Q_i^2 v_i^2  \nonumber \\ &&+ M_s^2 v_s - \lambda A_\lambda v_u v_d  \\
0&=& \tilde \lambda (\lambda v_s + \tilde \lambda v_{s^\prime}){v_u^2 + v_d^2 \over 2 }+ {{g^\prime}^2 Q_s^2 \over 2 }  \sum_i Q_i^2 v_i^2  \nonumber \\ &&+ M_{s^\prime }^2 v_{s^\prime} - \tilde \lambda A_{\tilde\lambda} v_u v_d 
\end{eqnarray}
where $\bar g = \sqrt{g_1^2 + g_2^2 }$, $v_u = v\sin\beta $ and $v_d =v\cos \beta$ with $v=246$ GeV from precision measurements. 
Given the tadpole conditions, one might write down the mass matrix of the CP-even Higgs in the basis $({\rm Re} H_u, ~{\rm Re} H_d,~ {\rm Re} S,~ {\rm Re}S^\prime)$, which takes the form
\begin{eqnarray}
\left( \matrix{M_{11}^2 & M_{12}^2 & M_{13}^2 & M_{14}^2  \cr  \spadesuit & M_{22}^2 & M_{23}^2 & M_{24}^2 \cr \spadesuit & \spadesuit & M_{33}^2 & M_{34}^2 \cr \spadesuit & \spadesuit&\spadesuit & M_{44}^2 }\right) \label{cpeven}
\end{eqnarray}
with
\begin{eqnarray}
M_{11}^2 &=&{v_u^2 \over 4 } (\bar g^2  + 4 g^{\prime 2 } Q_u^4)  +{ v_u \over v_d  } (\lambda A_{\lambda} v_s + \tilde \lambda A_{\tilde \lambda } v_{s^\prime }) \nonumber \\
M_{22}^2 &=& {v_d^2 \over 4 } (\bar g^2  + 4 g^{\prime 2 } Q_d^4)  +{ v_d \over v_u  } (\lambda A_{\lambda} v_s + \tilde \lambda A_{\tilde \lambda } v_{s^\prime }) \nonumber \\
M_{33}^2 &=& {\lambda A_\lambda v_u v_d -\lambda \tilde\lambda v^2 v_{s^\prime} \over 2 v_s } + g^{\prime 2 } Q_s^4 v_s^2  \nonumber \\
M_{44}^2 &=&{\tilde \lambda A_{\tilde \lambda} v_u v_d -\lambda \tilde\lambda v^2 v_{s} \over 2 v_{s^\prime} } + g^{\prime 2 } Q_s^4 v_{s^\prime}^2   \nonumber \\
M_{12}^2&=&\left[g^{\prime 2 } Q_u^2 Q_d^2 + \lambda^2+\tilde \lambda^2 -{\bar g^2 \over 4} \right] v_u v_d \nonumber \\&& - (\lambda A_\lambda v_s +\tilde \lambda A_{\tilde \lambda } v_{s^\prime} ) \nonumber \\
M_{13}^2 &=& v_u \left[g^{\prime 2 } Q_s^2 Q_u^2 v_s + \lambda (\lambda v_s +\tilde \lambda v_{s^\prime}) \right] -v_d \lambda A_{\lambda} \nonumber \\
M_{14}^2 &=&v_u \left[g^{\prime 2 } Q_s^2 Q_u^2 v_{s^\prime} + \tilde\lambda (\lambda v_s +\tilde \lambda v_{s^\prime}) \right] -v_d \tilde\lambda A_{\tilde\lambda} \nonumber \\
M_{23}^2 &=& v_d \left[ g^{\prime 2 } Q_d^2 Q_s^2 v_s + \lambda (\lambda v_s + \tilde \lambda v_{s^\prime }) \right] - v_u\lambda A_{\lambda } \nonumber \\
M_{24}^2 &=&  v_d \left[ g^{\prime 2 } Q_d^2 Q_s^2 v_{s^\prime} + \tilde \lambda (\lambda v_s + \tilde \lambda v_{s^\prime }) \right] - v_u\tilde \lambda A_{\tilde \lambda } \nonumber \\
M_{34}^2 &=& g^{\prime 2} Q_s^4 v_s v_{s^\prime } + {1\over 2 } \lambda \tilde \lambda v^2  \; .\nonumber 
\end{eqnarray}
The mass matrix in Eq. (\ref{cpeven}) can be diagonalized by either the $4\times 4$ unitary transformation or approximate blog diagonalization.  
In this paper we assume $\tilde \lambda$ is tiny, such that $S^\prime$ mainly mix with the $S$. 
As will be seen in the next section, ${\rm Re} S^\prime$ can explain the 750 GeV diphoton excess observed at the run-2 LHC. 

There is mixing between $Z$ and $Z^\prime$ arising from loop corrections and kinetic mixing. 
Phenomenological constraints require the mixing angle to be less than ${\cal O} (10^{-3})$~\cite{King:2005jy}. 
So the mixing can be neglected. 
The mass eigenvalue of the $Z^\prime$ can be written as $M_{Z^\prime} \approx  g^\prime Q_s \sqrt{v_s^2 +v_{s^\prime}^2}$. The exotic quark masses arise from eq. (\ref{exotic}), $M_{D_i^{} } = (\kappa_{2i} v_{s^\prime } + \kappa_{3 i } v_s)/\sqrt{2}$.  The mass matrix of the superpartners of the exotic quarks come from the F-terms and soft supersymmetry breaking terms, which, in the interaction eigenbasis $(\tilde D_i,~\tilde{\bar D_i})$, can be written as
\begin{eqnarray}
\left(  \matrix{ M_{D_i^{}}^2 + M_{\tilde D_i^1}^2 & A^{i} M_{D_i} - {1 \over 2 }v_u v_d (\lambda \kappa_3 +\tilde \lambda \kappa_2 )  \cr \spadesuit  & M_{D_i^{}}^2 + M_{\tilde D_i^2}^2 }\right)
\end{eqnarray} 
where $M_{\tilde D_i^1}^2 $ is the soft mass and $A^{i}$ is the trilinear coupling from soft SUSY breaking Lagrangian. 

The neutralino mass matrix, in the interaction basis $(\tilde B, ~\tilde W,  ~\tilde H_d^0, ~\tilde H_u^0 ,~\tilde S, ~\tilde S^\prime ,~\tilde B^\prime)$, reads
\begin{eqnarray}
\left( 
\matrix{ M_1 & 0 & -{1\over 2 } g^\prime v_d & {1\over 2 } g^\prime  v_u & 0&0&0 \cr  
                  \spadesuit  & M_2 & {1\over 2 } g v_d  & -{1\over 2 } g v_u & 0&0 &0 \cr
                   \spadesuit & \spadesuit & 0 & -\mu_E & - {\lambda v_u^{} \over \sqrt{2}} & -{\tilde\lambda v_u \over \sqrt{2}}& \tilde Q_d g^\prime v_d^{} \cr
                    \spadesuit& \spadesuit & \spadesuit & 0 & -{\lambda v_d \over \sqrt{2}} & -{\tilde \lambda v_d \over \sqrt{{2}}} & \tilde Q_u g^\prime v_u^{} \cr 
                     \spadesuit & \spadesuit & \spadesuit &\spadesuit & 0  & 0 & \tilde Q_s g^\prime v_s \cr
                        \spadesuit & \spadesuit & \spadesuit &\spadesuit & \spadesuit  & 0 & \tilde Q_s g^\prime v_{s^\prime} \cr 
                         \spadesuit & \spadesuit & \spadesuit &\spadesuit & \spadesuit  & \spadesuit & M_1^\prime } 
\right) \nonumber 
\end{eqnarray}
where $\mu_E= (\lambda v_s + \tilde \lambda v_{s^\prime}) /\sqrt{2}$, $M_1,~M_2$ and $M_1^\prime$ are the soft gaugino masses of $\tilde B,~\tilde W,~\tilde B^\prime$ respectively. 
The top-left $4\times 4$ blog  is the neutralino mass matrix of the MSSM by replacing $\mu$ with $\mu_E$.
The chargino mass matrix is the same as  that in the MSSM.

\begin{figure}[t]
  \includegraphics[width=0.45\textwidth]{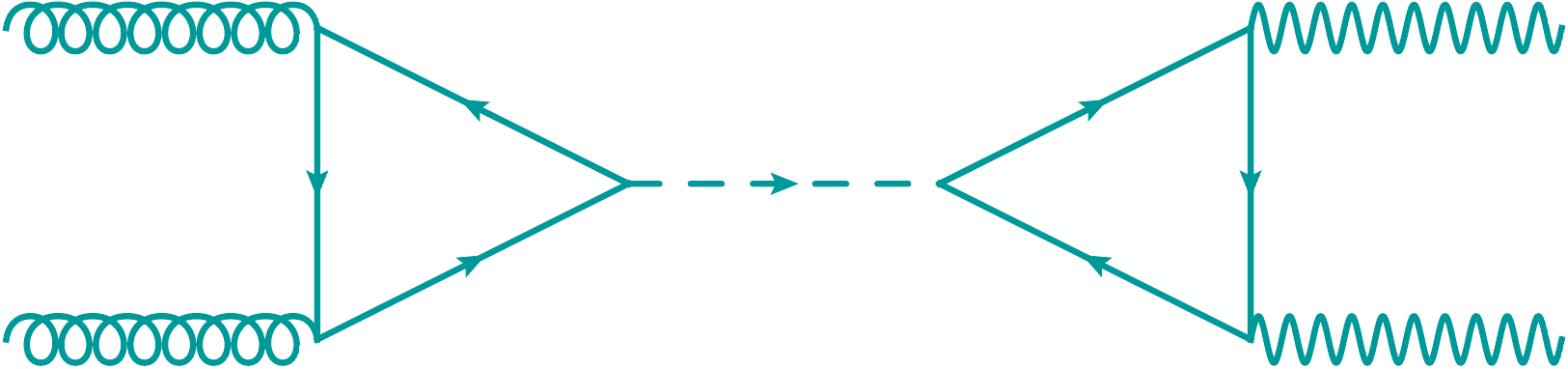}
\caption{\label{feyn}  Feynman diagram for the production of the resonance at the LHC.}
\end{figure}

\section{Diphoton excess}

Both the ATLAS and CMS collaborations have observed the diphoton excess at $m_{\gamma \gamma } =750~{\rm GeV}$, with the cross section roughly estimated as $\sigma(pp\to R \to \gamma \gamma) \approx 5\sim 10~{\rm fb}$.
According to Landau-Yang theorem~\cite{Landau:1948kw,Yang:1950rg}, this new resonance can only be spin-0 or spin-2 states. 
The gluon fusion or heavy flavor quark production of the resonance at the LHC are favored because the run-1 LHC did not see the excess clearly. 
In this section, we explain the diphoton excess as the signal of a scalar singlet in the ESSM. 
There are three fundamental scalar singlets in the ESSM,  two of them ($S$ and $S^\prime$) are responsible for the spontaneous breaking of the $U(1)_N$ in our model. 
They might be produced at the LHC through the gluon fusion with the exotic quarks in the loop and decay into diphoton mediated by the exotic quarks and Higgs-like particles.
The relevant feynman diagram is given in Fig.~\ref{feyn}. 
The cross section can be written as
\begin{eqnarray}
\sigma (pp\to R \to \gamma \gamma) ={ C_{gg}\over s M } \Gamma(R\to gg) {\rm BR} (R \to \gamma \gamma)
\end{eqnarray}
where $C_{gg}$ is the partonic integral, $\sqrt{s}$ is the center-of-mass energy, $M$ is the mass of the resonance.  
One has $C_{gg} \approx 3163$~\cite{Franceschini:2015kwy} at $\sqrt{s}=13~{\rm TeV}$.

\begin{figure}[t]
  \includegraphics[width=0.45\textwidth]{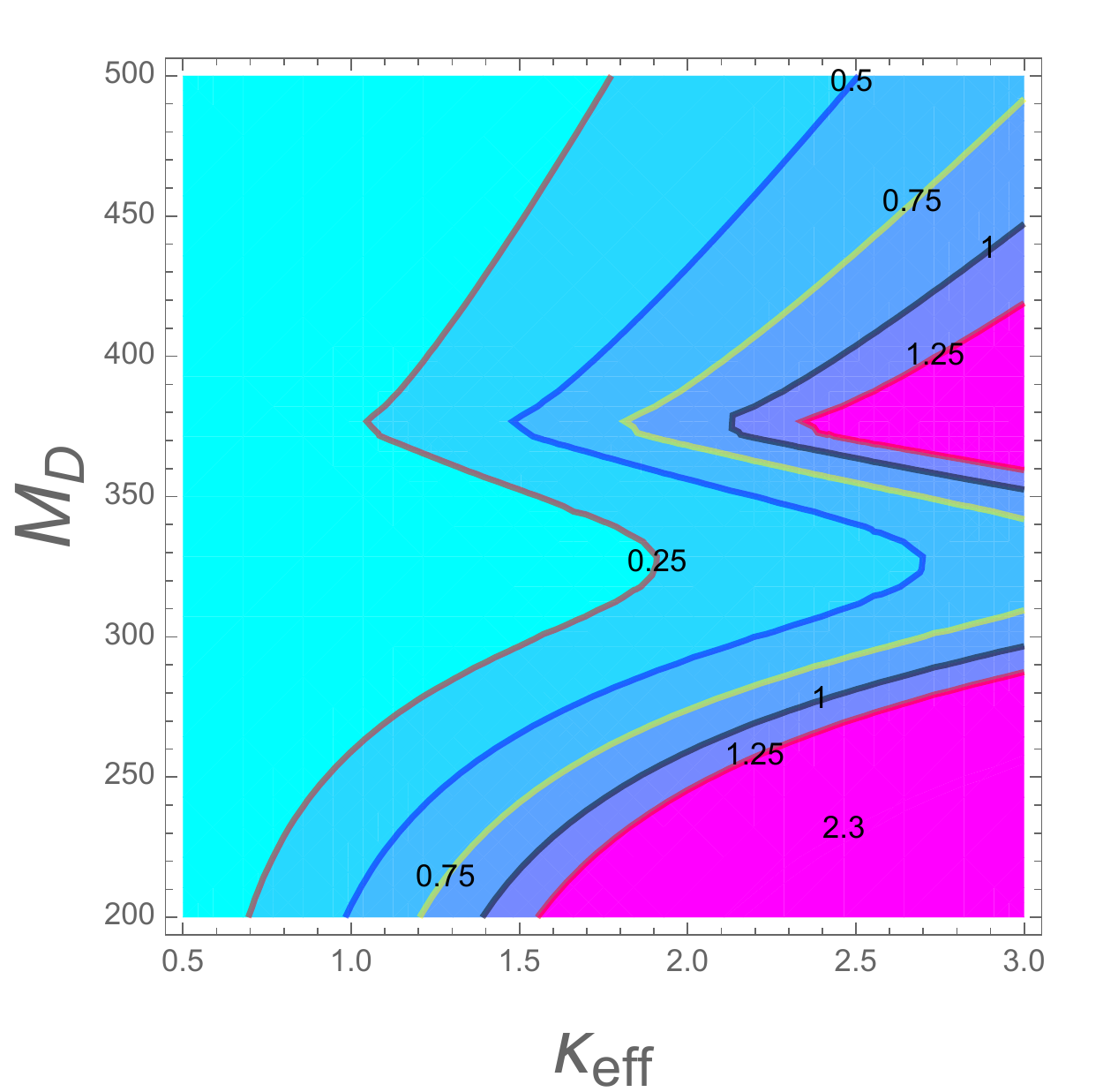}
\caption{\label{ggrate} Contours of the decay  rate of $\hat S^\prime \to gg $  in the $\kappa_{\rm eff}-M_D$ plane.}
\end{figure}

To fit the observed cross section, one needs to calculate the decay rates of the resonance to the digluon and diphoton. 
In the following study we assume the resonance is $\hat S^\prime$, which is the mass eigenstate of $S^\prime$.
As can be seen in the last section, both diquarks and their superpartners contribute to $\Gamma (S^\prime \to  gg )$. The diquarks mediated loops dominate the contribution, while the contribution of scalar loops are suppressed by the mixing angle that diagonalizes the  mass matrix in eq. (15). 
The decay rate can then be written as
\begin{eqnarray}
\Gamma_{gg} \approx { \alpha_s^2 M^3 \over 128 \pi^3 } \left|\sum_i^3 {\alpha \kappa_{2i}\over M_{D_i}} f(\tau )\right|^2
\end{eqnarray}
where $\alpha = \langle \hat S^\prime | S^\prime \rangle $, $\tau= 4 m_{D_i}^2 /M^2 $, $\alpha_s=g_s^2/4\pi$, and the function  $f(x)$ takes the form~\cite{Carena:2012xa}
\begin{eqnarray}
f(x) = -2x+2x(x-1)\arcsin^2 \left[{1\over \sqrt{x}} \right] \; ,\nonumber 
\end{eqnarray}
for $x>1$.  
Similarly one can get the expression of $\Gamma(\hat S^\prime \to \gamma \gamma)$.
Both the diquarks and the Higgs-like particles contribute to this process.
Again the scalar contribution is suppressed by the mixing angle.
The decay rate takes the form
\begin{eqnarray}
\Gamma_{\gamma \gamma } \approx {\alpha_e^2 M^3\over 1024 \pi^3 } \left| \sum_i^3 \left({2 \over 3 } {\alpha \kappa_{2i } \over M_{D_i}}  f(\tau_1) +{2 \alpha \lambda_{2i} \over M_{\tilde H_i }} f(\tau_{2})\right)\right|^2 
\end{eqnarray}
where $\tau_2 \equiv 4m_{\tilde H}^2/M^2$.

We show in Fig.~\ref{ggrate} contours of  $\Gamma_{gg}$ in the $\kappa_{\rm eff}-M_D$ plane, where we have assumed that the exotic quarks are degenerate with mass $M_D$ and couple to $\hat S^\prime$ with a universal effective coupling $\kappa_{\rm eff} \equiv \alpha \kappa_{2i}$.  
Apparently, one needs a strongly coupled regime to derive a relatively large $\Gamma_{gg}^{}$. 
By setting $\Gamma_{gg}=0.75~{\rm GeV}$, which correspond to $\kappa_{\rm eff}>1.8$ for $M_{D} >375~{\rm GeV}$, the production cross section of $\hat S^\prime$ at the run-2 LHC is  about $7.28$ pb. 
Notice that $M_{D} \ll M/2$ is not favored, because the total rate of $\hat S^\prime$ will be too large in this case.

\begin{figure}[t]
  \includegraphics[width=0.45\textwidth]{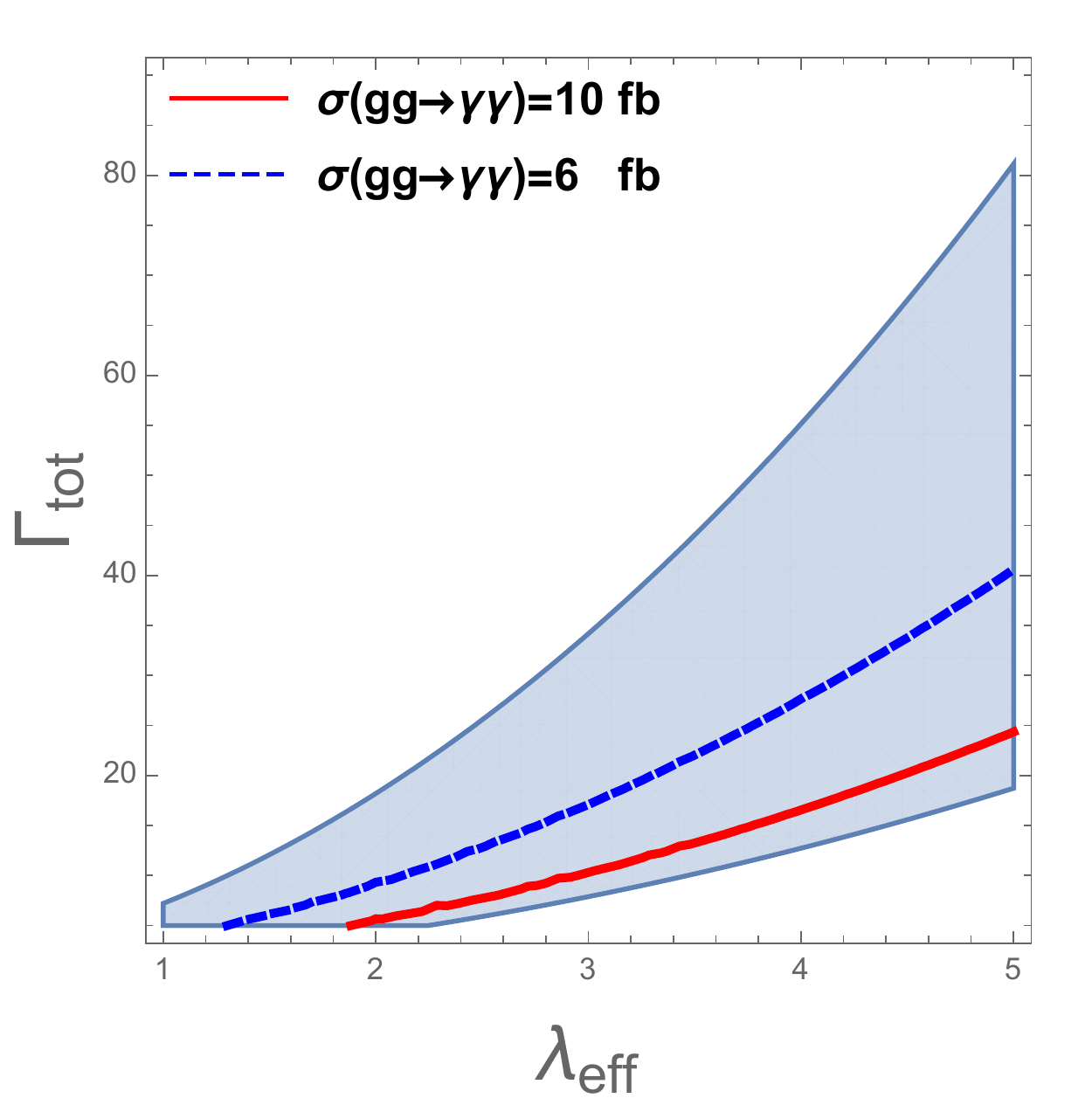}
\caption{\label{money} The region in the $\lambda_{\rm eff} - \Gamma_{\rm tot}$ plane that has $\sigma(gg\to\gamma \gamma) \in [3,~13]~{\rm fb}$.}
\end{figure}

We show in Fig.~\ref{money} the region in the $\lambda_{\rm eff} - \Gamma_{\rm tot}$ plane that has $\sigma(gg\to\gamma \gamma) \in [3,~13]~{\rm fb}$, by further assuming superpartners of Higgs-like particles are degenerate and have universal coupling, $\lambda_{\rm eff}$, with $\hat S^\prime$. 
The solid and dashed lines correspond to $\sigma(gg\to \gamma \gamma ) =10~{\rm fb}$ and $6~{\rm fb}$ respectively. 
We have set $\kappa_{\rm eff}=2$ and  $ M_D=M_{\tilde H} =380~{\rm GeV}$ when making the plot.  
The best fit width of the resonance is about $45~{\rm GeV}$. 
In this model, one needs $ \lambda_{\rm eff}\sim 4$ to get the best-fit width, which is very unnatural. 
As a conclusion, the model can hardly fit with the suggested width of the  resonance.
If the width is confirmed to be wide in the future, the model could be ruled out. 
In other words, if the ESSM could explain the diphoton excess, one needs to figure out the decay channels and the total rate of the resonance. In ESSM it might decay into superpartners of exotic quarks, $H_u^{1,2}$ and $H_d^{1,2}$ as well as SM particles through the mixing with the SM Higgs. 
We leave the systematic study of the decays  of $\hat S^\prime$ to a longer paper.

Finally let us check constraints on the model from the run-1 LHC at $\sqrt{s}=8~{\rm TeV}$. 
The upper bounds at the $95\%$ confidence level on cross sections of various final states produced from a resonance with $M=750$ GeV and $\Gamma=45$ GeV, are given as $\sigma\cdot {\rm BR } (ZZ)<12~{\rm fb}$~\cite{Aad:2015kna},  $\sigma\cdot {\rm BR} (hh)<39~{\rm fb}$~\cite{Aad:2015xja}, $\sigma\cdot {\rm BR} (t\bar t) <550~{\rm fb}$~\cite{Chatrchyan:2013lca} and $\sigma \cdot {\rm BR} (jj)<2.5~{\rm pb}$~\cite{Aad:2014aqa}. 
The first three cross sections put constraint on the mixing between the new resonance and the SM Higgs, which was already studied in many references. 
It corresponds to a small $\tilde \lambda$ in our scenario.
Here we only check the constraint of the dijet searches, which gives a upper limit to the ratio $r\equiv \Gamma_{gg}/\Gamma_{\gamma \gamma}$ and has $r<1300$~\cite{Franceschini:2015kwy}. 
For our case, $r \in (50,~400)$ as $\lambda_{\rm eff}$ varies from $1$ to $5$ for $\kappa_{\rm eff}=2$ and $M_D=M_{\tilde H} =380~{\rm GeV}$, which are the benchmark inputs of Fig.~\ref{money}.  
Thus this constraint is satisfied in this model.

\section{Conclusion}

An excess in the diphoton channel with invariant mass $m_{\gamma \gamma } =750~{\rm GeV}$ has been observed by the ATLAS and CMS collaborations. 
If confirmed, it will be a hint of new physics beyond the SM.  
In this paper we explain this resonance as the scalar singlet from a string-inspired exceptional supersymmetric standard model. 
The model we studied contains a $U(1)_N$ gauge symmetry, of which right-handed neutrinos carry no charge, and three electroweak singlets. 
The singlets couple to the exotic quarks and Higgs-like supermultiplets which are contained in the fundamental representation of the $E_6$, resulting in effective couplings of the singlets with diphoton and gluon-pair by integrating out new particles.    
Although both fermions and their superpartners contribute to the effective couplings, we claimed that the scalar contributions, which are suppressed by the mixing angles diagonalizing scalar mass matrices, are negligible.
The model turns out might produce the observed cross section, but can hardly fit with the wide width of the resonance. 
We expect the future running of the LHC shows us more clear hints of the resonance, which might approve or rule out our model.

\begin{acknowledgments}

The author thanks to Huai-ke Guo, Ran Huo, Hao-lin Li, Grigory Ovanesyan  and Jiang-hao Yu for very helpful discussions. 
This work was supported in part by DOE Grant DE-SC0011095.

\end{acknowledgments}

\end{document}